\documentclass[journal=jctcce,manuscript=article]{achemso}
\usepackage[version=3]{mhchem} 
\usepackage{setspace}
\usepackage{amsmath}
\usepackage{amssymb}
\usepackage{mathtools}
\usepackage{amsthm}
\usepackage{graphicx}
\usepackage{multirow}
\usepackage{amsfonts}
\usepackage{bbm}
\usepackage{caption}
\usepackage{graphicx}
\usepackage{dcolumn}
\usepackage{bm}
\usepackage{enumitem}
\usepackage{etoolbox}
\usepackage{microtype}
\usepackage{subcaption}
\usepackage{booktabs} 
\usepackage{xcolor}
\usepackage{achemso}
\usepackage{soul}

\newcommand{\vecx}{\textbf{x}}

\newcommand{\vece}{\textbf{e}}

\theoremstyle{plain}

\theoremstyle{definition}

\theoremstyle{remark}

\author{Wenqi ZENG}
\affiliation[math]
{Department of Mathematics, The Hong Kong University of Science and Technology, Clear Water Bay, Kowloon, Hong Kong SAR 999077, China}

\author{Lu ZHANG}
\affiliation[cas]{State Key Laboratory of Structural Chemistry, Fujian Institute of Research on the Structure of Matter, Chinese Academy of Sciences, Fuzhou, Fujian 350002, China}

\author{Yuan YAO}
\affiliation[math]
{Department of Mathematics, The Hong Kong University of Science and Technology, Clear Water Bay, Kowloon, Hong Kong SAR 999077, China}
\alsoaffiliation{Department of Chemical and Biological Engineering, The Hong Kong University of Science and Technology, Clear Water Bay, Kowloon, Hong Kong SAR 999077, China}
\email{yuany@ust.hk}

 \title{Leveraging Transformer Models to Capture Multi-Scale Dynamics in Biomolecules by nano-GPT}

\begin{document}

\begin{tocentry}

    \centering
    \includegraphics[width=0.8\linewidth]{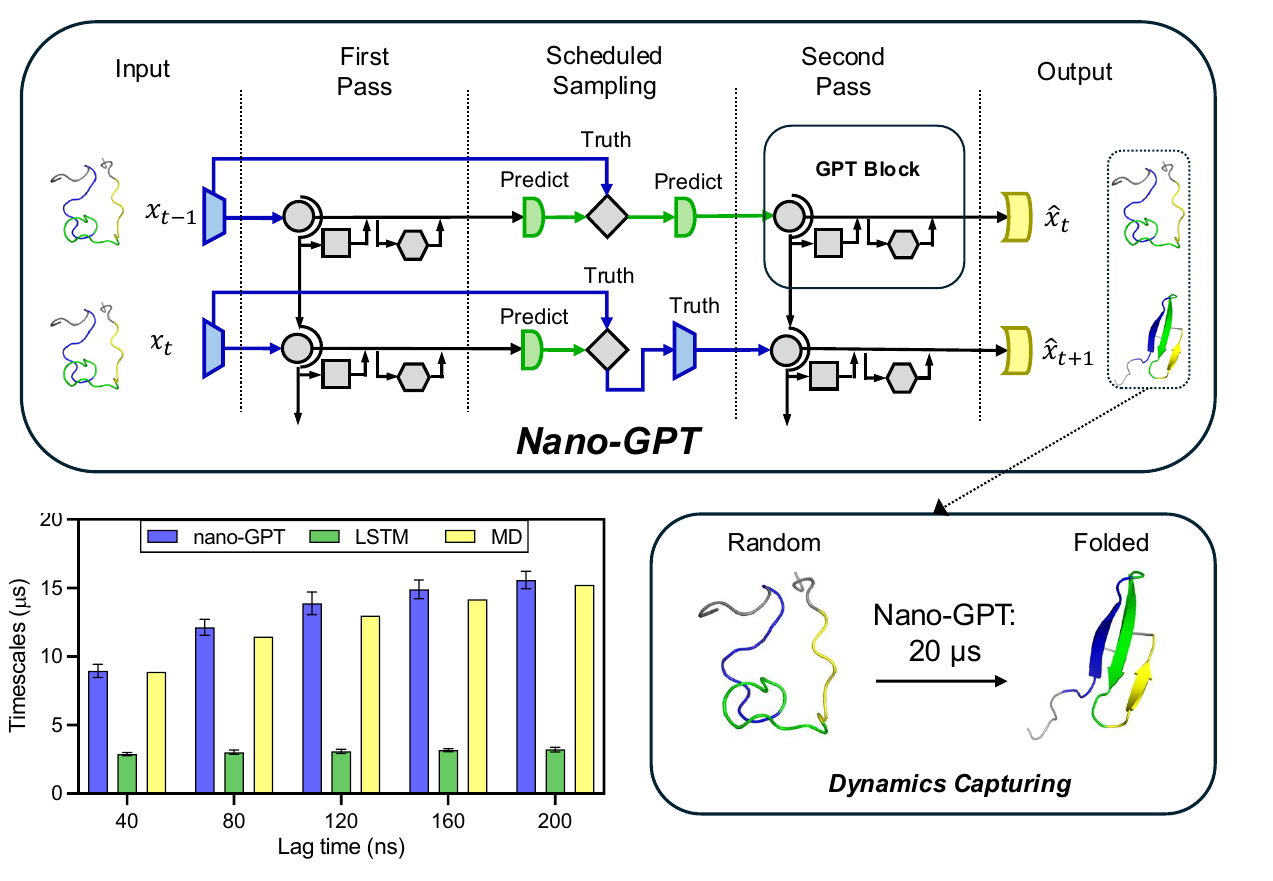}






\end{tocentry}

\begin{abstract}
Long-term biomolecular dynamics are critical for understanding key evolutionary transformations in molecular systems. However, capturing these processes requires extended simulation timescales that often exceed the practical limits of conventional models. To address this, shorter simulations, initialized with diverse perturbations, are commonly used to sample phase space and explore a wide range of behaviors. Recent advances have leveraged language models to infer long-term behavior from short trajectories, but methods such as long short-term memory (LSTM) networks are constrained to low-dimensional reaction coordinates, limiting their applicability to complex systems. In this work, we present nano-GPT, a novel deep learning model inspired by the GPT architecture, specifically designed to capture long-term dynamics in molecular systems with fine-grained conformational states and complex transitions. The model employs a two-pass training mechanism that incrementally replaces molecular dynamics (MD) tokens with model-generated predictions, effectively mitigating accumulation errors inherent in the training window. We validate nano-GPT on three distinct systems: a four-state model potential, the alanine dipeptide, a well-studied simple molecule, and the Fip35 WW domain, a complex biomolecular system. Our results show that nano-GPT effectively captures long-timescale dynamics by learning high-order dependencies through attention mechanism, offering a novel perspective for interpreting biomolecular processes.

\end{abstract}

\section{Introduction}
\label{Introduction}


Biomolecular dynamics play a crucial role in understanding the function of biological macromolecules. These dynamics encompass a wide range of conformational changes, from short-term and local vibrations to long-term and significant conformational shifts, occurring over various time scales. They lay the structural basis for facilitating processes such as enzyme catalysis, molecular recognition, and signal transduction, thus critical to the proper functioning of proteins, nucleic acids, and other biomolecules \citep{leimkuhler1996integration}. Therefore, accurately predicting the multi-scale conformational changes, especially the long-timescale dynamics, can advance our understanding about the structure-activity relationship of biomolecules, and guide the enzyme engineering and drug design.

Molecular dynamics (MD) simulations are a widely used tool to investigate the conformations of biomolecules, and provides the dynamics for localized and high-frequency events occurring at nanoseconds to a few microseconds. However, the critical conformational changes for functioning are usually rare events, involving the transitions between different conformational states occurring at timescales from microseconds to milliseconds or even longer. Such a long-timescale dynamics is challenging to reach directly by conventional all-atomic MD simulations \citep{dullweber1997symplectic}. The timescale gap between the short-term MD simulations and long-term dynamics hinders a full understanding of the conformational dynamics of the macromolecules as well as their functions. In this regard, extracting the long-term pattern from short simulations is critical to address the challenges in predicting biomolecular dynamics \citep{chodera2007automatic, pan2008building, noe2013variational}.

One commonly used approach for extracting long-term dynamics from short simulations is Markov State Models (MSMs) \citep{husic2018markov}, which are effective for systems with well-defined, slow dynamics and clear timescale separation. However, the construction of accurate MSMs requires careful selection of the lag time and states. An inappropriate choice of lag time can miss fast dynamics or overly smooth slow processes, while poor state discretization can obscure meaningful transitions. Additionally, MSMs are less suited for modeling fast, hidden, or non-Markovian processes, which are increasingly common in complex molecular systems. In contrast, Long Short-Term Memory (LSTM) networks \citep{gers2000learning} excel at modeling sequential data without requiring explicit selection of predefined states or lag times, unlike methods such as MSMs and HMMs \citep{husic2018markov, bowman2009progress}. LSTMs can learn complex, high-order Markovian dynamics directly from the data and generate new frames, making them particularly well-suited for modeling conformational transitions that span multiple timescales \citep{tsai2020learning, tsai2022path}. \cite{tsai2020learning} shows that LSTM accurately capture Boltzmann statistics and kinetics across systems like alanine dipeptide and riboswitches. As a result, LSTMs offer a more efficient and flexible alternative to MSMs for biomolecular simulations. Building on previous work, \citet{tsai2022path} tackled the challenge of learning longer dynamics using even shorter frames to train the LSTM. By incorporating static and dynamic constraints within an iterative framework guided by the Maximum Caliber principle, they successfully predicted 2 ns of transition dynamics from simulations as short as 0.2 ps. However, these methods primarily address simplified systems with coarse-grained, low-dimensional projections (e.g., $\phi$ and $\psi$ in alanine dipeptide). In complex systems, LSTMs struggle to capture long-range dependencies, such as oversmoothing transitions in all-atom alanine dipeptide or missing the slowest dynamics in Fip35.

The difficulties LSTMs encounter in capturing long-range dependencies can be traced to inherent limitations in their sequential processing structure \citep{zhao2020rnn, dieng2017topicrnn}. Although LSTMs are designed to retain and update information across time steps through gating mechanisms, the standard initialization of these gates can hinder the learning of long-term temporal correlations \citep{tallec2018can}. Over time, the information stored in the LSTM’s memory cell tends to decay exponentially, restricting the network’s ability to maintain relevant information from distant time steps \citep{gu2020improving}. Consequently, LSTMs are often less effective at capturing slow or rare dynamics that require the retention of information over extended periods. Generative Pre-trained Transformer (GPT)-like models \citep{brown2020language} address the long-range dependency limitations of LSTMs by utilizing parallel processing through matrix operations. The self-attention mechanism within GPT enables the model to capture more intricate relationships between distant frames in the sequence, without relying on strict sequential processing \citep{taylong}. Recent work by \citep{bera2025accurate} applies a decoder-only Transformer to predict future states in molecular systems. While effective on systems with limited state space and continuous trajectory inputs, their approach does not address the challenges of modeling long-range dependencies in de novo sequence generation or in systems with finer-grained state discretization.

In this paper, we introduce nano-GPT, a novel method specifically designed to capture long-timescale molecular dynamics from short frames of unbiased MD simulations. By learning non-Markovian dependencies, nano-GPT generates states to extend the original trajectory, demonstrating the time evolution of states. Theoretically, we establish a connection between state embeddings in nano-GPT and kinetic time in dynamics. Experimentally, we validate nano-GPT on three systems of varying complexity: a model potential, alanine dipeptide, and the Fip35 WW domain. Nano-GPT captures both statistical and dynamic features across complex systems and low-dimensional simplified systems, excelling in metrics such as free energy and mean first passage time (MFPT). Our study offers a GPT-based method to predict the dynamics of complex system, nano-GPT can effectively capture critical information across distant frames, overcoming key limitations of traditional methods like LSTM.

\section{Methods} \label{sec:method}

\begin{figure}[H]
\includegraphics[width=0.7\linewidth]{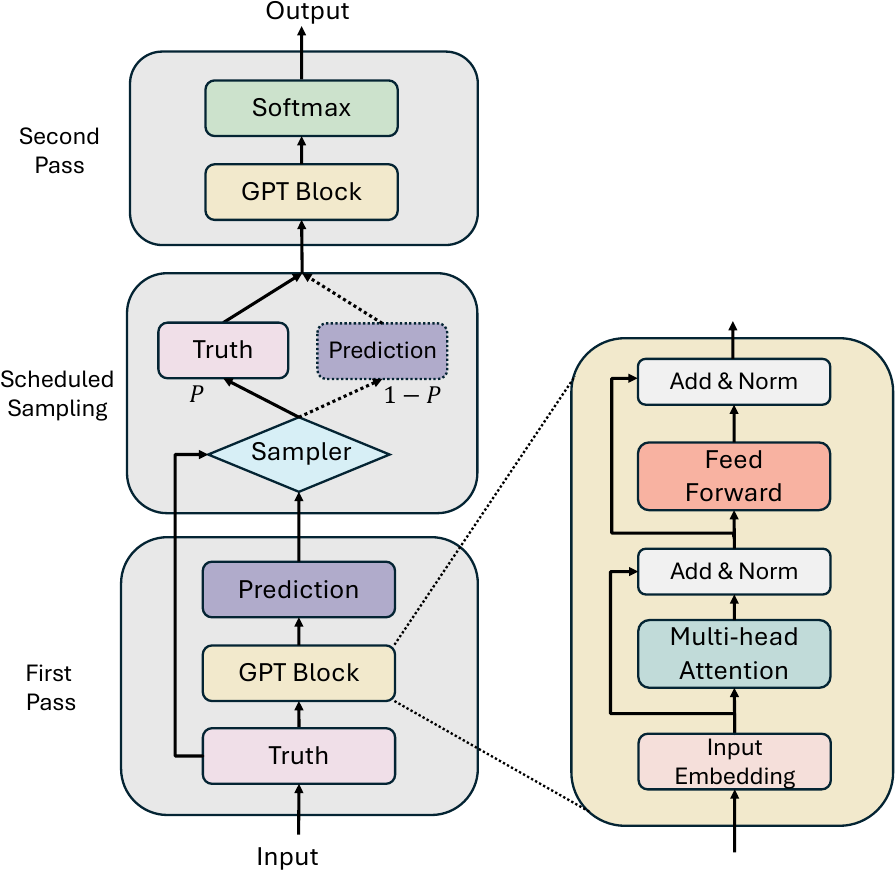}
\caption{Model structure of Nano-GPT. It consists of two sequential passes, linked by a sampler. The first pass works as a standard decoder, where ground truth tokens (from MD simulations) are provided as input to generate initial predictions. A sampler then selects tokens either from these predictions or the ground truth, based on a dynamically changing probability, and forwards them to the second pass. In the second pass, the sampled tokens are used to generate the final output.}
\label{fig:model}
\vspace{0.2in}
\end{figure}

This section describes the workflow of the nano-GPT model for predicting molecular dynamics (MD) sequences. As illustrated in Fig.~\ref{fig:model}, nano-GPT incorporates three key phases to learn molecular dynamics. In the \textit{first pass}, a GPT block transforms the input sequence into embeddings that capture the kinetic time between states, which is crucial for understanding metastable dynamics. Next, during the \textit{scheduled sampling}, a sampler is employed to mitigate the biases present in short MD simulations. The sampler dynamically adjusts the balance between using ground truth MD tokens and model-generated predictions during training, improving the model’s ability to generalize. Finally, the \textit{second pass}, including another GPT block, processes the sampled tokens from the sampler, generating the model’s final output. The following sections detail the model design and theoretically prove that training with cross-entropy loss optimizes the system’s path entropy, assuming first-order Markovianity and ergodicity.

In our experiment, we employ a sliding window of length 50 to incorporate a larger portion of the data. For the simpler dataset, 4-state system, alanine dipeptide on $\psi$ and $\phi$ coordinates, we use a learning rate of 0.0005 and two GPT blocks. For the more complex datasets, including alanine dipeptide based on RMSD and the Fip35 WW protein, the learning rate is increased to 0.001, with three GPT blocks utilized. Across all datasets, the hidden dimension is fixed at 128, with a batch size of 32. For the LSTM model used as comparison, a learning rate of 0.001 is used, with 512 recurrent units and the stateful parameter set to True. These hyperparameters, consistent across all datasets, were identified as optimal through random hyperparameter search. Both nano-GPT and LSTM models are intentionally lightweight, enabling efficient execution on a 2080Ti GPU. 

To prepare the data for training, we concatenate trajectories sequentially into a single long sequence of states. The sequences are simply appended one after another without introducing artificial transitions or markers between them. Although this results in a continuous state sequence, the model is not biased by this artificial continuity due to the subsequent processing: the concatenated sequence is divided into overlapping windows of a fixed length (as training window size specified in Supporting Information), and these windows are randomly shuffled during training. This ensures that the model learns generalizable temporal patterns rather than memorizing the order of concatenated trajectories. In next part, we first outline the base GPT architecture, the foundation of our model, followed by a comprehensive analysis of the two-pass learning mechanism and sampling strategy. Our code for all systems can be found on \url{https://github.com/Wendysigh/MD_nano_GPT}.



\paragraph{GPT model with attention mechanism:} Given a sequence of discrete MD states, $[x_1, ..., x_t]$, GPT is designed to predict the subsequent state, $\hat{x}_{t+1}$, and generate additional samples. The raw sequence $[x_1, ... , x_t] $ will be transformed to embeddings as $\textbf{X} := [\vecx_1, ..., \vecx_t]^T \in \mathbb{R}^{t \times d_{\rm x}}$, where $d_{\rm x}$ is the embedding dimension. Within the GPT model, the embeddings $\textbf{X}$ are transformed into hidden vectors $\textbf{H}^{(l)}$ at the $l$-th layer. The model then processes $\textbf{H}^{(l)}$ to produce the final probability distribution $\mathrm{Q}(x_{t+1} \mid x_{\leq t})$, which predicts the next element in the sequence. A pivotal component of the GPT architecture is the self-attention mechanism, which transforms the input embeddings $\textbf{X}$ into hidden vectors $\textbf{H}^{(l)}$ through following steps:

First, the $\textbf{X}$ is projected into query vectors $\textbf{Q} \in \mathbb{R}^{d_{\rm q} \times d_{\rm x}}$, key vectors $ \textbf{K} \in \mathbb{R}^{d_{\rm q} \times d_{\rm x}}$, value vectors $\textbf{V} \in \mathbb{R}^{d_{\rm q} \times d_{\rm x}}$. By initializing $\textbf{H}^{(0)} = \textbf{X}$, 
The output for next layer $\textbf{H}^{(l)}$ is then defined as:
\begin{align}
    \textbf{A}^{(l)} &= softmax(\frac{\textbf{Q}^{(l)}(\textbf{K}^{(l)})^T}{\sqrt{d_{\rm q}}}) \textbf{V}^{(l)}, \label{eq:A}\\
    \textbf{B}^{(l)} &= f_{\bm{\theta}}(\textbf{H}^{(l-1)} + \textbf{A}^{(l)} ) \label{eq:B}\\
    \textbf{H} ^{(l)} &=  \textbf{H}^{(l-1)} + \textbf{A}^{(l)} + \textbf{B}^{(l)}, \label{eq: H}
\end{align}

In Eq.~\ref{eq:A}, $\textbf{A}$ is also known as attention matrix, which stands for the contextual information learned for every token in input sequence. 
We use $f_{\bm{\theta}}$ to denote the non-linear transformation in Eq.~\ref{eq:B}, standing for the feed forward network in Fig.~\ref{fig:model}. This network comprises a two-layer neural architecture followed by a normalizing nonlinear operator.

The decoding steps that yield the final probability distribution $\mathrm{Q}(x_{t+1} \mid x_{\leq t})$ again with a non-linear transformation $f_{\bm{\theta}}$, involving layer normalization and a residual connection. Following $f_{\bm{\theta}}$, the model employs a linear projection, represented by $\textbf{D} \in \mathbb{R}^{d_{\rm out} \times d_{\rm x}}$, and concludes with a softmax activation function. This sequence of operations effectively transforms the hidden representations into a probability distribution over potential output tokens.

\begin{equation}\label{eq:condi_prob}
    \mathrm{Q}(x_{t+1} \mid x_{\leq t}) = softmax (f_\theta (\textbf{H}^{(l)}) \textbf{D}^T + \textbf{b} )
\end{equation}

\paragraph{Within Pass: token embedding captures kinetic time.} 
Eq.~\ref{eq:condi_prob} can be rewritten as follow, where $\vece_m \in \mathbb{R}^{d_{\rm out}}$ is a one-hot vector with the m-th element non-zero.  
\begin{equation}\label{eq:condi_prob2}
    \mathrm{Q}(x_{t+1} = m \mid x_{\leq t}) = \frac{\exp((f_\theta (\textbf{H}^{(l)}) \textbf{D}^T + \textbf{b} ) \times \vece_m)} 
{\sum_k \exp((f_\theta (\textbf{H}^{(l)}) \textbf{D}^T + \textbf{b} ) \times \vece_k)}
\end{equation}

By using Taylor's theorem, the $f_\theta (\textbf{H}^{(l)})$) can be approximated around a differentiable point $\textbf{X} = \textbf{m}$:
\begin{equation*}
    f_\theta (\textbf{H}^{(l)}) \approx f_\theta (\textbf{H}^{(l)})|_{\textbf{X} = \textbf{m}} + (\textbf{X} - \textbf{m}) \textbf{M}_\theta^T
\end{equation*}
where $\textbf{M}_\theta$ is defined as $(\textbf{M}_\theta )_{ij} = \frac{\partial (f_\theta)_i}{\partial \vecx_j}|_{\textbf{X}=\textbf{m}}$.

Recall $\textbf{H}^{(0)}$ is initialized as $\textbf{X}$, $\textbf{H}^{(l)}$ can be rewritten as:
\begin{equation}\label{eq:H_sum}
    \textbf{H}^{(l)} = \textbf{X} + \sum_{k=0}^l (\textbf{A}^{(k)} + \textbf{B}^{(k)})
\end{equation}

By Eq.~\ref{eq:H_sum}, Eq.~\ref{eq:condi_prob2} becomes,
\begin{equation}\label{simple_prob}
        \mathrm{Q}(x_{t+1}= m \mid x_{\leq t}) = \frac{\exp(\textbf{C}_m) \exp( \textbf{X} \textbf{M}_\theta^T \textbf{D}^T \vece_m)} 
{\sum_k \exp(\textbf{C}_k) \exp( \textbf{X} \textbf{M}_\theta^T \textbf{D}^T \vece_k)}
\end{equation}
where $\textbf{C}_m = [f_\theta (\textbf{X} + \sum_{k=0}^l (\textbf{A}^{(k)} + \textbf{B}^{(k)}))|_{\textbf{X} = \textbf{m}}  -  \textbf{m}\textbf{M}_\theta^T )\textbf{D}^T + \textbf{b} ] \times \vece_m$. 

In Eq.~\ref{simple_prob}, $\textbf{M}_\theta^T \textbf{D}^T \vece_m$ can be treated as the output embedding for m-th state with the projection matrix as $\textbf{M}_\theta^T \textbf{D}^T$, noted as $\hat{\textbf{X}} ^{(m)} := \textbf{M}_\theta^T \textbf{D}^T \vece_m $. Similarly to \citep{tsai2020learning}, $\textbf{C}_m$ is a correction term for the time lag effect. While there is no exact calculation for such correction term, under first order Markovian assumption, the transition probability between two states $\mathrm{Q}_{ml} \coloneqq \mathrm{Q}(x_{t+1} = m \mid x_t=l)$ can be rewritten as a ansatz: 
\begin{equation} \label{eq: emb dis}
    \mathrm{Q}_{ml} = \frac{ \exp(\textbf{C}_m) \exp(\textbf{X}^{(l)}  \cdot \hat{\textbf{X}} ^{(m)} )}
    { \sum_k \exp(\textbf{C}_k) \exp(\textbf{X} ^{(l)}  \cdot \hat{\textbf{X}} ^{(k)} )}
\end{equation}



\begin{equation}
    \label{eq: kinetic time}
    t_{lm} = \frac{1}{P_l * \mathrm{Q}_{ml} + P_m * \mathrm{Q}_{lm}}
\end{equation}

The kinetic time defined in Eq.~\ref{eq: kinetic time}, or equivalently average transition time, can be measured as the inverse of interconversion probability, where $P_l$ stands for the Boltzmann distribution calculated for state $l$.

In other words, the model embeddings hold information for kinetic time. In the experiments section, we demonstrate that these embeddings contain sufficient information to accurately recover the final prediction, suggesting their importance in capturing dynamical relations.


The ansatz in Eq.~\ref{eq: emb dis} requires the input sequence with state $l$ and state $m$ in subsequent position. However, such a requirement will be impractical if the simulations fail to provide such coverage. Furthermore, short MD simulations may contain noisy irrelevant perturbation and causing distractions to discern subtle long-term dynamics. To address these issue, we propose the scheduled sampling technique in next part.

\paragraph{Across Pass: sampler mitigates short simulation bias.} As shown in Figure~\ref{fig:model}, our model utilizes scheduled sampling between the two-pass training, gradually substituting golden tokens with its own predictions. 


In detail, the first forward pass operates as a standard decoder, outputting weighted sums of target embeddings as probabilities. The second forward pass uses sampled tokens, chosen from either the golden tokens or the first-pass predictions. The sampling probability follows a decaying scheme based on the $i$-th training step and $t$-th decoding position.


\begin{equation}
    p =
    \left\{
        \begin{array}{cc}
                \epsilon ^ {t (1 - k^i)} & \mathrm{choose \ golden \ token} \\
                1-\epsilon ^ {t (1 - k^i)} & \mathrm{otherwise}\\
        \end{array} 
    \right.
\end{equation}

where $\epsilon$ and $k$ are constants in the range $(0, 1)$. This scheme ensures that for smaller training steps and decoding positions, the model is exposed to more ground-truth tokens. As training progresses and decoding advances, the model increasingly relies on its own outputs during training. This gradual shift is guided by a composite exponential decay schedule, adapted from \cite{liu2021scheduled}, which effectively balances supervision with self-generated prediction. A visualization illustrating the effects of different composite strategies, along with a table summarizing the hyperparameters, is provided in the Supplementary Information (SI).


The scheduled sampling takes effect in three key ways: (i) Noisy short simulations are increasingly replaced by the model's predictions as optimization progresses, enhancing model comprehension in later stages. (ii) During training, models consistently receive the correct previous token as input. However, during generation, models rely on their own
previously generated tokens. Scheduled sampling narrows this gap between training (using ground-truth data) and inference (relying on their own outputs). (iii) Cross-entropy loss, designed for single-label classification, optimizes by maximizing the probability of the single observed next state. It does not explicitly model the uncertainty over multiple plausible next states that could have occurred \citep{liu2021scheduled}. Scheduled sampling addresses this by introducing more diverse input sequences, improving the model’s ability to handle transition variability. Nevertheless, this limitation does not contradict our earlier proof that minimizing cross-entropy is equivalent to maximizing path entropy when the data distribution is ergodic and sufficiently covers all plausible transitions. In that case, the recovery of transition uncertainty arises from the ergodicity of the data, not from the properties of the loss function itself.




\paragraph{Cross-entropy minimization encourages path entropy maximization:} 
The optimization objective is the cross-entropy loss , as defined in Eq.~\ref{eq:loss}, calculated over the entire sequence. Here, $\mathrm{P}(x_{t+1} \mid x_{\leq t})$ denotes the true distribution, while $\mathrm{Q}(x_{t+1} \mid x_{\leq t})$ denotes the predicted distribution for $x_{t+1}$. 

\begin{equation}
    \ell = -\sum_{t=0}^{T}  \sum_{x_{t+1}} \mathrm{P}(x_{t+1}   \mid x_{\leq t}) \cdot \ln \mathrm{Q}(x_{t+1}  \mid x_{\leq t}) 
    \label{eq:loss} 
\end{equation}

Under the framework of Maximum Caliber~\citep{presse2013principles}, path entropy $J$ is defined in Eq.~\ref{eq:path entropy}, where $\mathrm{Q}_{ml} \coloneqq \mathrm{Q}(x_{t+1} = m \mid x_t=l)$ , $\mathrm{P}_{ml} \coloneqq \mathrm{P}(x_{t+1} = m \mid x_t=l)$, $\mathrm{P}_{l} \coloneqq \mathrm{P}(x_{t+1} = l)$. 

\begin{equation}
    \mathit{J} = -T \sum_{lm} \mathrm{P}_l \mathrm{P}_{ml} \cdot \ln \mathrm{Q}_{ml}
    \label{eq:path entropy}
\end{equation}

\begin{equation}
    \mathit{J} = - \sum_{t=0}^T \sum_{m} \mathrm{P}_{ml} \cdot \ln \mathrm{Q}_{ml}
    \label{eq:path entropy rewrite}
\end{equation}

The key to the proof lies in treating Eq.~\ref{eq:path entropy} as an ensemble average for $\sum_{m} \mathrm{P}_{ml} \cdot \ln \mathrm{Q}_{ml}$. For a large enough $T$ and assuming ergodicity, the ensemble average can be replace by the time average as shown in Eq.~\ref{eq:path entropy rewrite}. Under the assumptions of first-order Markovianity and ergodicity, Eq.~\ref{eq:path entropy rewrite} can be directly obtained from Eq.~\ref{eq:loss}.

The proof established by \citep{tsai2020learning} for LSTM models is directly applicable to nano-GPT. Therefore, we do not provide an extensive proof in this paper, as the conclusions can be readily extended to our model.



%


\section{Results and Discussions} \label{sec:experiment}

Built on a GPT-like architecture and enhanced with scheduled sampling \citep{pang2021text}, nano-GPT is a lightweight two-pass GPT-based model, as shown in Fig.~\ref{fig:model}.  Across the two passes, a scheduled sampler progressively replaces MD simulation tokens with predictions, reducing bias from short simulations and improving long-term forecasting accuracy. The performance of nano-GPT is evaluated on three systems: a 4-state model potential, alanine dipeptide, and the Fip35 WW domain. Nano-GPT outperforms LSTM by more accurately capturing long-timescale dynamics and aligning with MD ground truth in metrics like free energy, ITS, and MFPT. In alanine dipeptide, nano-GPT closely matches the first ITS and MFPT, while LSTM overestimates these values. In Fip35, nano-GPT captures the slowest dynamics, including the longest ITS of 14 µs and MFPT of 18 µs, while LSTM underestimates both, demonstrating nano-GPT’s superior performance in modeling complex molecular systems.


The experiments across the three systems span a range of timescales and challenges related to slow dynamics, which are evaluated using ITS and MFPT—larger values indicate slower transitions. These dynamics are primarily influenced by two factors: (1) training window size, which determines the temporal span of input sequences. A longer window provides the model with more global temporal context; (2) number of discrete states, which controls the granularity of conformational space discretization. A higher number of states provides finer state resolution, leading to lower equilibrium probabilities per state and splitting transitions across additional intermediate states, which may introduce noise or make the model sensitive to small fluctuations rather than important state transitions. We provide an ablation experiment of the effect of number of states and training window in SI.

The challenges of each system are as follows: ${\rm Alanine}_{\psi}$ has a 10 ps training window, 20 states, and slowest dynamics around 150 ps; ${\rm Alanine}_{\phi}$ has a 10 ps training window, 20 states, and slowest dynamics around 30 ns; ${\rm Alanine}_{\rm RMSD}$ has a 20 ps training window, 100 states, and slowest dynamics around 80 ns; and the Fip35 WW domain has a 20 ns training window, 100 states, and folding dynamics around 18 µs. As the system complexity increases from alanine to the Fip35 WW domain, capturing long-term behaviors becomes more challenging, leading to an increase in prediction difficulty. 

\begin{figure*}[ht]
  \centering
{\includegraphics[width=0.95\linewidth]{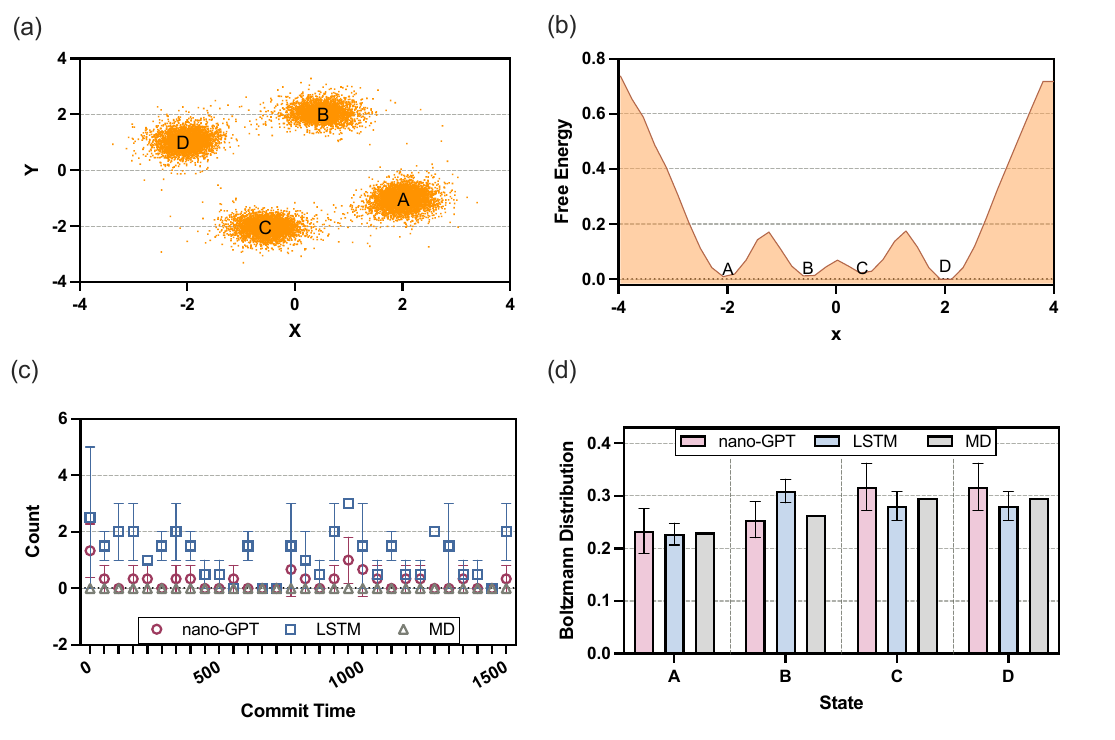}}
\caption{Performance in 4-state across (a) free energy landscape. (b) projection on x-axis (c) transition count from state A to state D, as a function of commit time as defined in ``Results" section. (d) Boltzmann distribution. The error bars represent the standard deviation computed over 3 independent simulations for both LSTM and nano-GPT.}
\label{fig:4-states}
\vspace{0.2in}
\end{figure*}




We compare the performance of nano-GPT and LSTM \citep{tsai2020learning} in capturing the long-term molecular dynamics from the MD ground truth, where performance closer to MD is better. Three key metrics are used, Free Energy, Implied Time Scales (ITS) and Mean First-Passage Time (MFPT). The free energy represents the thermodynamic potential of a system and helps to describe the stability of different molecular conformations. Implied Time Scales (ITS) quantifies the characteristic timescales over which a system transitions between different states, reflecting both slow and fast processes. A long ITS indicates that the system remains in a particular state for an extended period, suggesting stability or metastability, while a short ITS indicates rapid transitions between states. In the experiments, we focus on the 1st and 2nd ITS, corresponding to the slowest and second slowest dynamic behaviors. The mean first-passage time (MFPT) quantifies the average time it takes for a system to transition from one state to a target state for the first time, providing insight into the ease or difficulty of such transitions. A long MFPT suggests that the system is slow to escape from an initial state, often due to energy barriers, while a short MFPT indicates rapid transitions. Together, these metrics offer a comprehensive view of the system’s kinetic behavior, helping to reveal both long-term stability and the timescales of rare or slow events.




\subsubsection{4-state System}
The potential energy landscape for the four-state system, which represents a symmetric system with four discrete metastable states, is constructed using the methodology outlined by \citet{tsai2020learning}. The data along x-coordinate is mapped to the nearest of four predefined states, simplifying the system while retaining key dynamics. This experiment evaluates nano-GPT and LSTM models in replicating the system’s transition behaviors and equilibrium distributions under high energy barriers, a hallmark of molecular dynamics. 

The free energy landscape, shown in Fig.\ref{fig:4-states} (a), reveals high energy barriers separating states A and D. Such barriers significantly hinder state transitions, leading to inefficient sampling in MD simulations. As a result, the system tends to remain trapped in local energy minima for extended periods, with transitions between states being rare. This behavior is reflected in Fig.\ref{fig:4-states} (b), which depicts the transition count as a function of \textit{commit time}, which refers to the minimum duration a system must remain in a state before a subsequent transition is considered irreversible. In MD simulations, the transition count approaches zero for substantial commit times, indicating inefficient sampling.

Despite the challenges posed by the energy landscape, nano-GPT successfully replicates the transition dynamics observed in MD, accurately modeling the near-zero transition counts for high commit times. In contrast, LSTM deviates more noticeably from the ground truth. This difference arises from the architectural strengths of the models. nano-GPT, with its transformer-based architecture and self-attention mechanisms, excels at capturing long-term dependencies and rare transitions. It can leverage global sequence context to model the sparse and infrequent transitions characteristic of high energy barriers. On the other hand, LSTM relies on localized memory updates and gating mechanisms, which are less effective for handling long-range dependencies and rare events, leading to its reduced accuracy in this aspect.

The equilibrium distribution of the system, shown in Fig.\ref{fig:4-states} (c), provides additional insights. This distribution represents the steady-state probabilities of the system. Both nano-GPT and LSTM produce equilibrium distributions with acceptable fluctuations, demonstrating their ability to capture the steady-state properties of the relatively simple four-state system. Since the equilibrium distribution does not heavily depend on long-term temporal correlations, both models perform adequately in this context. However, the more accurate replication of transition dynamics by nano-GPT suggests its superior capability in handling both equilibrium and dynamic aspects of the system.

In summary, 4-state experiment demonstrates the effectiveness of nano-GPT in modeling molecular systems with high energy barriers and sparse transitions. While LSTM performs reasonably well in capturing equilibrium distributions, its limitations in handling long-term dependencies and rare transitions make it less effective in accurately modeling transition dynamics. nano-GPT’s transformer-based architecture provides a distinct advantage, enabling it to excel in scenarios where global sequence context and long-term temporal patterns are critical. This highlights the potential of transformer models like nano-GPT in advancing the simulation and analysis of complex molecular systems.

\subsubsection{Alanine Dipeptide}
The alanine dipeptide, comprising 22 atoms and 66 Cartesian coordinates, is a benchmark model to assess the capability of the kinetic model in predicting the conformational dynamics. The dataset consists of 100 trajectories, each including the dipeptide and 888 water molecules, with atomic positions recorded at 0.1 ps intervals over a total duration of 1 µs. We analyze the alanine dipeptide system using $\psi$ (psi) and $\phi$ (phi) as reaction coordinates, representing local conformational changes, and RMSD as a global structural metric to assess overall structural deviations.


The $\psi$ angle serves as a reaction coordinate that primarily captures faster conformational motions, providing insight into the model’s ability to resolve short-timescale fluctuations accurately. The $\phi$ angle, on the other hand, exhibits slower dynamics, making it a benchmark for evaluating the model’s performance on long-timescale processes and its capacity to preserve memory of the system’s evolution. The RMSD is a global structural metric that integrates information across the entire molecular conformation, capturing the combined effects of both fast and slow dynamics. Additionally, it introduces noise in the whole phase space, offering a challenging test of the model’s robustness and ability to distinguish signal from noise.

By incorporating these three settings, we ensure a rigorous evaluation of the model’s performance across diverse dynamical regimes, from localized fast transitions to global structural changes, and its ability to capture both timescale-dependent and spatially integrated features.

\begin{figure*}[ht]
   \centering
    \includegraphics[width=1\linewidth]{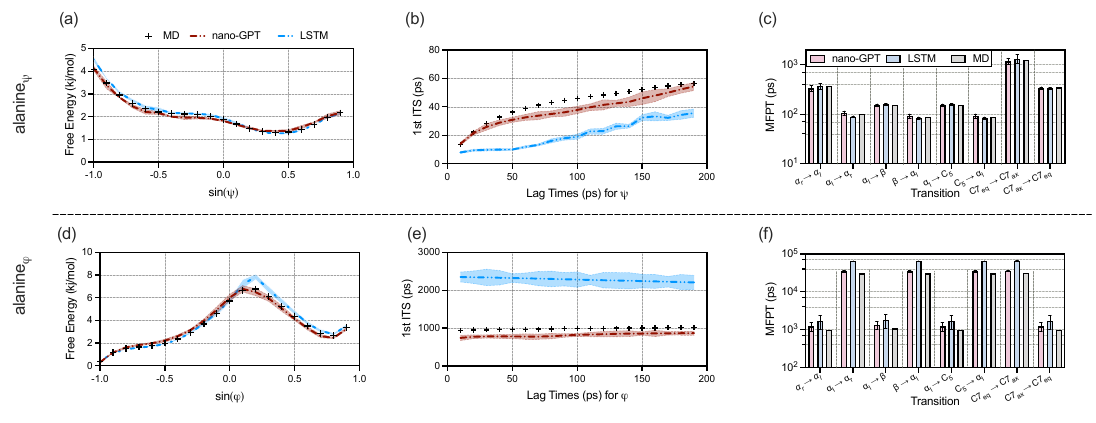}
    \caption{Results on alanine dipeptide for $\psi$ (upper row) and $\phi$ (lower row): Performance comparison across free energy landscape, ITS, and MFPT. (a) Free energy landscape, (b) ITS, and (c) MFPT for ${\rm alanine}{\psi}$; (d) Free energy landscape, (e) ITS, and (f) MFPT for ${\rm alanine}{\phi}$. Shaded areas in (a)(b)(d)(e) and error bars in (c)(f) represent the standard deviation computed over 3 independent simulations for both LSTM and nano-GPT. The ITS and MFPT are calculated with a lagtime of 10ps on the two-dimensional dihedral space, where the lag time corresponds to the discrete time interval between observations in the trajectory used for computing transition probabilities or identifying state transitions.}
\label{fig:all}
\vspace{0.2in}
\end{figure*}


\paragraph{Fast motions in torsion angle $\psi$}: 
The $\psi$ coordinate plays a critical role in capturing and analyzing the fast dynamics of alanine dipeptide. In this context, the original molecular trajectories are projected onto the torsional angle $\psi$, where the majority of degrees of freedom are associated with rapid motions, such as the vibrations of chemical bonds. Such projection simplifies the high-resolution molecular system into a lower-dimensional representation that retains the essential features of fast dynamical processes. Under this setting, we attempt to use a short training window of 10 ps to predict the dynamics over 150 ps.

\begin{figure}[t]
   \centering
\includegraphics[width=\textwidth]{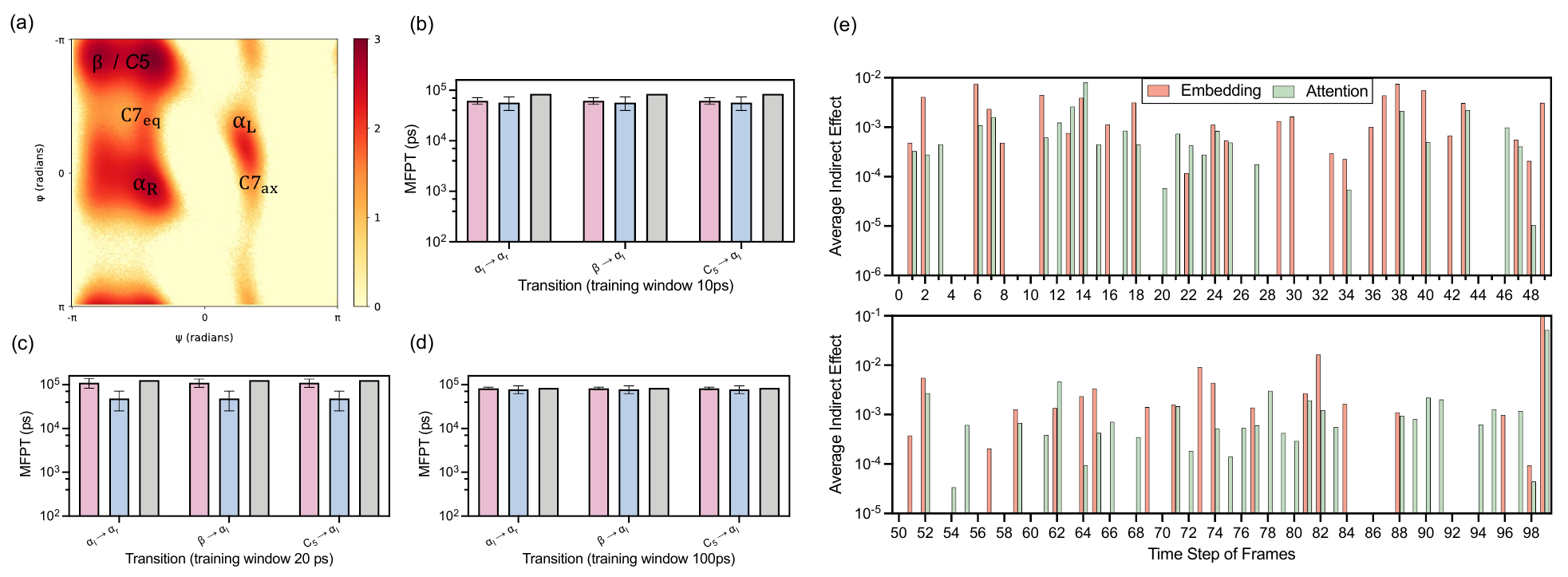}
\caption{ Analysis on $\rm{Alanine}_{\rm RMSD}: $ (a) Ramachandran plot of alanine dipeptide: four metastable states ($\beta$, $\alpha_R$, $\alpha_L$, C7ax)
with $\psi$ on the vertical axis and $\phi$ on the horizontal axis. The metastable states are located at:
$\beta$ (C7eq) in the top-left, $\alpha_R$ (alpha helix) in the left-center, $\alpha_L$ (left-hand helix) in the right-center, and C7ax in the bottom-right. (b)(c)(d) Comparison of MFPT (ps) with different training window (10 ps, 20 ps \& 100 ps). Values closer to the MD ground truth reflect better performance. (e) Influence of embeddings and attention mechanisms across 100 frames on the final prediction. Higher average indirect effect (AIE) indicates greater importance. Smaller time step of frame reflects longer-term dependency in the prediction.}
\label{RMSD}
\end{figure}

Fig.~\ref{RMSD} (a) illustrates the metastable states in the Ramachandran plot, showing clear separation between dominant conformations. Specifically, the $\psi$ angle captures transitions between key conformational states, such as the $\alpha_R$ and $C7ex$ states. These transitions predominantly occur on shorter timescales, reflecting the fast dynamics inherent to the torsional degrees of freedom along $\psi$.


Although relatively easy, nano-GPT still performs better than LSTM on the ${\rm alanine}_{\psi}$. 
Fig.~\ref{fig:all} (a) plots the free energy landscape in ${\rm alanine}_{\psi}$, which reflects free energy landscape include both thermodynamics and kinetics. Both nano-GPT and LSTM accurately replicates the true curve and trend as the MD ground truth. Overall, the 1st ITS presents the most challenging dynamic behavior in ${\rm alanine}_{\psi}$. In terms of ITS in Fig.~\ref{fig:all} (b), the 1st ITS captured by nano-GPT (red line) closely matches the ground truth MD (green line), while LSTM predicts a biased and mistakenly fast ITS (blue line). For the 2nd ITS, which represents shorter timescales and is easier to predict, both models align well with the ground truth. Regarding Mean First-Passage Time (MFPT) in Fig.~\ref{fig:all} (c), both nano-GPT and LSTM exhibit consistency with the ground truth, particularly in predicting the slowest mode of 1265 ps from $C7eq$ to $C7ax$.

$\psi$ is a crucial coordinate for evaluating the model’s ability to resolve fast conformational changes. Furthermore, it provides a direct measure of how well the model captures the kinetic barriers and rapid transitions between metastable states, which are essential for understanding the short-timescale dynamics of the system. By focusing on $\psi$, nano-GPT can assess the fine-grained temporal resolution of the model in a computationally efficient and interpretable manner.

\paragraph{Slow motions in torsion angle $\phi$}: The $\phi$ coordinate provides a complementary perspective to $\psi$. Unlike $\psi$, which predominantly reflects fast conformational changes, dynamics along $\phi$ exhibit more intricate transitions. These transitions often involve interactions between competing energetic basins, such as those linking the $\alpha_R$ and $C7ex$ states or subtle variations within the $\alpha_R$ basin itself. This intricacy arises due to the stronger coupling of $\phi$ with slower, larger-scale structural rearrangements, such as backbone rotations and steric interactions. Transitions along $\phi$ often influence, and are influenced by, global structural properties, making it more representative of backbone flexibility and steric constraints. This coupling makes $\phi$ dynamics more complex and reflective of the molecule’s overall structural evolution. Under this setting, we attempt to use a short training window of 10 ps to predict the dynamics over 30 ns.


Nano-GPT demonstrates superior performance compared to LSTM across all three evaluation metrics — free energy landscape, ITS, and MFPT — by closely aligning with the molecular dynamics (MD) baseline. In contrast, the LSTM model consistently overestimates these metrics and exhibits incorrect behavior, particularly in the context of slower motions and rare transitions. 

In the free energy landscape (Fig.~\ref{fig:all} (d)), Nano-GPT accurately reproduces MD-like high-energy regions, which are critical for capturing subtle, low-probability conformational states. Conversely, LSTM deviates in these regions, suggesting an over-simplified representation of the potential energy surface. This overestimation likely stems from LSTM’s inability to fully capture the nuanced interplay between local fluctuations and global structural transitions. 

Fig.~\ref{fig:all} (e) reveals significant discrepancies in the ITS values between the two models. Nano-GPT accurately predicts the 1st ITS at approximately 80 ps, reflecting the dominant slowest motion in the system, while the LSTM model overestimates this timescale, with fluctuations extending to 250 ps. Such overestimation indicates that LSTM may blur transitions between metastable states, leading to artificially prolonged timescales.

As shown in Fig.~\ref{fig:all} (f), nano-GPT excels in modeling rare transitions, with its MFPT values closely matching those observed in MD simulations. In contrast, LSTM overestimates MFPTs, particularly for infrequent events. This discrepancy suggests that LSTM might exaggerate the energy barriers separating metastable states, a common issue when models fail to correctly capture the interplay of fast and slow dynamics. 

By including $\phi$ in our analysis, we gain additional details regarding slower, more global dynamics that complement the fast, localized transitions captured by $\psi$. LSTM shows an overestimation of $\phi$ dynamics due to its recurrent structure, which relies on a sequential mechanism. While this mechanism is effective for capturing short-term dependencies, it may introduce bias when modeling complex, multi-scale dynamics. In contrast, Nano-GPT’s attention-based architecture enables it to dynamically focus on relevant features across all timescales, allowing for a more precise and flexible representation of molecular dynamics.




\paragraph{Whole phase space in $\textbf{Alanine Dipeptide}$}

For a global structural analysis, trajectories are directly decomposed into discrete states using the root mean square displacement (RMSD) distances without any preprocessing of the high-resolution MD data. The discrete states are derived using a $k$-center clustering method to split the conformation space \citep{zhao2013fast}, which approximates an $\epsilon$-cover of samples \citep{Sun2008, Yao_JCP09, Yao_JCP13} based on the RMSD distances of heavy atoms (non-hydrogen). ${\rm Alanine}_{\rm RMSD}$ characterizes the entire conformational space. These four slowest modes correspond to the following transitions: the transition from $\alpha_R$ to structures on the left side, the transition between $\alpha_L$ and C7eq, and the transitions between $\alpha_R$ and C7ax. Under this setting, we attempt to test short training windows of 10 ps, 20 ps, or 100 ps to predict the dynamics over an extended period of 80 ns.

Unlike simpler reaction coordinates such as $\phi$ and $\psi$, ${\rm alanine}_{\rm RMSD}$ is the most challenging dataset due to its global structural information and growing number of discrete states. RMSD characterizes the entire conformational space by capturing transitions between metastable states. These transitions reflect not only local changes but also global shifts in the molecule’s backbone structure, offering a comprehensive view of the system’s dynamics. Thus, it provides a rigorous benchmark for assessing model performance under challenging conditions and is ideal for evaluating a model’s capacity to handle large-scale complex datasets.

As shown in Fig.~\ref{RMSD}(c), nano-GPT demonstrates an advantage over LSTM in capturing long-term global dynamics, evidenced by a noticeable performance gap. In Fig.~\ref{RMSD}(b) and Fig.~\ref{RMSD}(d) where the overall results are comparable, however, nano-GPT exhibits smaller error bars, indicating more consistent and stable predictions across independent runs. The training window, which represents the temporal context available to the models, significantly affects their ability to predict molecular transitions accurately. For ${\rm alanine}_{\rm RMSD}$, Nano-GPT demonstrates remarkable efficiency, effectively learning long-term dynamics across 100 states with training windows as short as 10 ps or 20 ps. This is evident in accurately predicting mean first passage times (MFPTs) for key transitions such as $\alpha_L$ to $\alpha_R$, $\beta$ to $\alpha_L$, and $C_5$ to $\alpha_L$, which occur on much longer timescales of 80 ns to 100 ns. By leveraging its attention mechanism, Nano-GPT dynamically captures dependencies across the entire sequence, enabling it to replicate these transitions with high fidelity, as reflected in its predicted trajectories aligning closely with MD baselines.

In contrast, LSTM struggles to achieve the same level of accuracy in capturing slow dynamics between metastable states. At shorter training windows, such as 10 ps or 20 ps, LSTM fails to predict MFPTs reliably, reflecting its limited ability to integrate long-term dependencies. While extending the training window to 100 ps improves LSTM’s performance by providing more local information, this improvement is limited to specific transitions, such as those involving $\alpha_L$. The root of this limitation lies in LSTM’s sequential structure, which is inherently constrained in representing complex dependencies over extended timescales. Nano-GPT, by contrast, excels in handling both local and global dynamics due to its ability to directly model relationships between all input frames, making it better suited for capturing rare transitions and accurately reproducing long-term dynamics, even under constraints of shorter training windows.

To further investigate the internal mechanisms of Nano-GPT on ${\rm alanine}_{\rm RMSD}$, we highlight its ability to effectively learn long-term dependencies. By leveraging the causal trace technique from \citep{meng2022locating}, as detailed in the SI, we analyze the temporal influence of embeddings and attention mechanisms on the model’s predictions. This approach provides valuable insights into how Nano-GPT integrates information across time steps to make accurate predictions.

Fig.~\ref{RMSD}(c) presents an in-depth analysis of how Nano-GPT processes both local and global details through the average indirect effect (AIE), a metric that quantifies the influence of embeddings and attention mechanisms on the model’s final predictions. The experiment spans 100 frames, with the 99th frame being closest to the final prediction and the 0th frame representing the most distant point in time. Notably, a higher AIE value in the \textit{distant} frames reflects the model’s ability to incorporate \textit{long-term} information into its decision-making process. This finding highlights Nano-GPT’s exceptional capacity to integrate temporally distant yet contextually relevant information, underscoring its advantage in modeling the intricate long-term dynamics of molecular systems.

In Fig.~\ref{RMSD} (c), both embeddings and attention mechanisms exhibit the highest Average Indirect Effect (AIE) at the 99th frame, the frame closest to the final prediction. This high importance near the prediction point aligns with expectations, as embeddings and attention mechanisms naturally have a direct, short-term influence on the outcome. However, if the model were unable to capture long-term dependencies, the AIE values at earlier (distant) frames would drop to zero, reflecting a loss of relevant information over time. Remarkably, in nano-GPT, the AIE values for both embeddings and attention are distributed across all 100 frames, with significant contributions from distant frames, such as the 1st, 2nd, and 6th frames. This sustained influence from earlier time points demonstrates Nano-GPT’s robust capacity to integrate temporally distant information, a key factor in its ability to model and predict long-term molecular dynamics effectively.

Furthermore, the comparable AIE values between embeddings and attention mechanisms highlight the critical roles both components play in Nano-GPT’s decision-making process. The embeddings’ significance arises from their ability to encode essential molecular information, such as kinetic times for metastable states, as discussed in Eq.~\ref{eq: kinetic time}. By effectively encoding and preserving this information, embeddings contribute to the model’s accurate representation of molecular dynamics, complementing the attention mechanism’s role in capturing contextual relationships. Together, these components enable Nano-GPT to achieve superior performance in modeling complex molecular systems.

\subsubsection{Fip35 WW Domain}

\begin{figure}[t]
   \centering
\includegraphics[width=.95\textwidth]{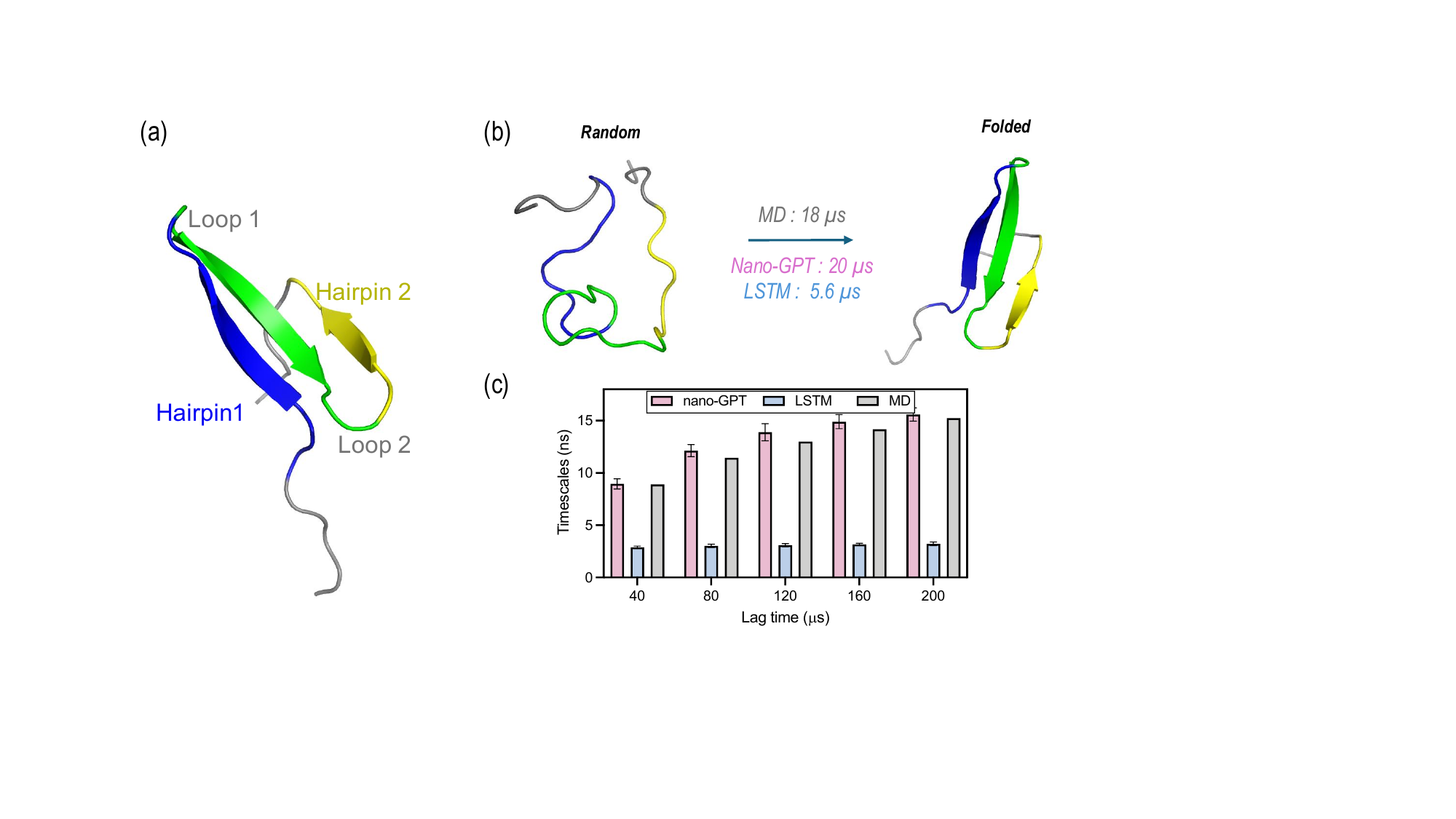}
\caption{Analysis on Fip35 WW domain, from left to right: (a) representative structure of Fip35 WW domain, (b) random coil structure, folded structure, (c) 1st ITS for different lag times. }
\label{FIP35}
\end{figure}


In the previous sections, we demonstrated that both nano-GPT and LSTM can learn and capture the slow dynamics of reaction coordinate $\psi$ and $\phi$. However, only nano-GPT successfully resolves the all-atom fluctuations in $\rm{alanine}_{\rm{RMSD}}$. While alanine dipeptide serves as a minimal model for studying basic torsional dynamics, the FiP35 WW domain introduces significantly greater challenges due to its complex folding mechanisms, diverse timescales, and multidimensional energy landscape.

The FiP35 WW domain is a 35-residue protein with a well-defined $\beta$-sheet structure. Modeling its folding involves capturing cooperative interactions such as hydrogen bonding, hydrophobic packing, and long-range contacts, which significantly increase the system’s complexity. Additionally, FiP35’s energy landscape is multidimensional and populated with numerous metastable states, making sampling rare transitions computationally intensive and requiring advanced techniques such as replica exchange or metadynamics. The Fip35WW dataset \citep{shaw2010atomic} is preprocessed using tICA \cite{perez2013identification} and k-center \citep{zhao2013fast} to convert raw MD trajectories into discrete states. Following previous study, tICA with three components and a lag time of 10 ns, combined with k-center clustering, is employed to construct the discrete state representation of the system. This representation is sampled at 0.2 ns intervals, covering a total simulation duration of 1.1 ms, providing a detailed and temporally resolved depiction of the system’s dynamics.



The results demonstrate a significant disparity in the ability of nano-GPT and LSTM to capture the slow dynamics of the FiP35 WW domain within the 20 ns training window. Nano-GPT accurately captures the longest ITS of 14 µs, closely aligning with the expected dynamics of the system, while LSTM underestimates the ITS, capturing a much faster dynamic at 3 µs. Similarly, mean first passage times (MFPT) from the random coil to the folded structure (typically around 20 µs) show that nano-GPT replicates this behavior with a value of 18 µs. In contrast, LSTM underestimates the MFPT, predicting much faster folding behavior. 

This marked difference in performance can be attributed to the fundamental architectural differences between the two models and the nature of the datasets. LSTM forgets or fails to integrate distant, critical information from earlier frames, resulting in an underestimate of the true ITS and MFPT. In alanine dipeptide, the system’s simplicity allows LSTM to perform relatively well despite its architectural limitations, sometimes overestimating dynamics due to its tendency to smooth out transitions in low-dimensional systems. However, the FiP35 WW domain’s high dimensionality and rugged energy landscape amplify LSTM’s shortcomings, as the model cannot effectively navigate the intricate conformational space or resolve subtle energy barriers. This causes LSTM to truncate slower transitions and focus disproportionately on faster, more apparent motions.

Nano-GPT’s attention mechanism allows it to dynamically integrate information across all input frames, regardless of their temporal distance. This flexibility enables it to accurately model both short-term fluctuations and long-term dependencies, crucial for resolving the hierarchical dynamics of FiP35 WW domain. Unlike LSTM, nano-GPT effectively captures the intricate interplay of local and global motions in the folding process, leading to accurate ITS and MFPT predictions that closely mirror the true system behavior.

The shift from overestimation in alanine dipeptide to underestimation in FiP35 WW domain for LSTM reflects its sensitivity to system complexity and timescale diversity. While LSTM can over-smooth and exaggerate simpler dynamics, it struggles to capture long-term dependencies and subtle energy barriers in more complex systems like FiP35, leading to an underestimation of dynamics. Nano-GPT, by leveraging its attention mechanism, overcomes these limitations, delivering consistently accurate predictions across both simple and complex systems.



\section{Conclusions}


Understanding biomolecular dynamics, especially long-timescale conformational changes, is critical for uncovering structure-activity relationships and advancing fields such as enzyme engineering and drug discovery. Traditional approaches like Markov State Models (MSMs) and Long Short-Term Memory (LSTM) networks face limitations in capturing complex, non-Markovian behavior or long-range dependencies. In contrast, Generative Pre-trained Transformer (GPT)-based models leverage self-attention to effectively model long-range, high-order dynamics in parallel. Building on this, our nano-GPT approach accurately predicts long-timescale behavior from short MD trajectories, capturing both statistical and kinetic properties across a range of systems. Evaluations on the 4-state potential, alanine dipeptide, and Fip35 WW domain show that nano-GPT outperforms existing methods in reproducing free energy landscapes and mean first passage times.

Future directions focus on improving the transferability and scalability of nano-GPT, especially under limited data conditions. One goal is to apply nano-GPT to larger and more complex biomolecular systems, such as multi-protein assemblies, to assess its robustness across a wider range of biological contexts. However, achieving reliable performance in such systems typically requires large amounts of training data and presents challenges in transferability. To address this, another key priority is to enhance the model’s ability to generalize by transferring learned representations across different molecular systems. Developing effective strategies for cross-system knowledge transfer would help reduce data demands and greatly expand nano-GPT’s practical utility for modeling biomolecular dynamics.

\newpage
\begin{acknowledgement}
We thank X. Huang, S. Cao for many useful discussions. This work was supported by the Research Grants Council (RGC) of Hong Kong, SAR, China (GRF-16308321) and National Natural Science Foundation of China (NSFC)(22273107). Y.Y. gratefully acknowledges the NSFC/RGC Joint Research Scheme Grant N\_HKUST635/20. Part of this work was carried out using the X-GPU cluster supported by the RGC Collaborative Research Fund C6021-19EF.

\end{acknowledgement}


\newpage





\begin{suppinfo}
Supporting Information:
\begin{itemize}
    \item Transformer layer architecture and structural details
    \item Hyperparameter specifications, training and simulation settings
    \item Experimental setting comparison across datasets
    \item Ablation studies: scheduled sampler, discrete state count, training window size
    \item Complete transition path visualization for 4-state model potential
    \item Causal trace analysis demonstrating attention effects in alanine dipeptide system
\end{itemize}

\end{suppinfo}


\clearpage 
\bibliography{ref}

\end{document}